\documentclass[aps,prl,twocolumn,a4paper,superscriptaddress,amsmath,amssymb,showpacs]{revtex4}
\usepackage[dvips]{graphicx}
\usepackage[dvips]{color} 
\usepackage{subfigure}

\renewcommand{\vec}[1]{\mathbf #1}
\renewcommand{\exp}{\mathrm e}

\renewcommand{\d}{\mathrm d}

\renewcommand{\hbar}{\hslash}

\begin{document}

\title{Elementary Charge Transfer Processes in a
  Superconductor-Ferromagnet Entangler}

\author{Jan Petter Morten}
\email{jan.morten@ntnu.no}
\author{Daniel Huertas-Hernando}
\affiliation{Department of Physics, Norwegian University of Science
  and Technology, N-7491 Trondheim, Norway}
\affiliation{Centre for
  Advanced Study, Drammensveien 78, N-0271 Oslo, Norway}
\author{Wolfgang Belzig}
\affiliation{University of Konstanz, Department of Physics, D-78457
  Konstanz, Germany}
\affiliation{Centre for
  Advanced Study, Drammensveien 78, N-0271 Oslo, Norway}
\author{Arne Brataas}
\affiliation{Department of Physics, Norwegian University of Science
  and Technology, N-7491 Trondheim, Norway}
\affiliation{Centre for
  Advanced Study, Drammensveien 78, N-0271 Oslo, Norway}

\date{August 17, 2007}

 \begin{abstract}
   We study the production of spatially separated entangled electrons
   in ferromagnetic leads from Cooper pairs in a superconducting lead.
   We give a complete description of the elementary charge transfer
   processes, i) transfer of Cooper pairs out of the superconductor by
   Andreev reflection and ii) distribution of the entangled
   quasiparticles among the ferromagnetic leads, in terms of their
   statistics. The probabilities that entangled electrons flow into
   spatially separated leads are completely determined by
   experimentally measurable conductances and polarizations. Finally,
   we investigate how currents, noise and cross correlations are
   affected by transport of entangled electrons.
 \end{abstract}
 
  \keywords{}

\maketitle

A solid state entangler is an electronic analog of the optical setups
used for experimental Bell inequality tests, quantum cryptography and
quantum teleportation \cite{tittel:qic01}.  Ideally, such a device
should produce separated currents of entangled electrons.
Superconductors are suitable candidates as sources in solid state
entanglers since Cooper pairs constitute entangled states.  This
prospect has motivated several papers addressing the properties of
hybrid superconductor and normal metal entanglers
\cite{NS-entanglers,entanglerfilters,entanglerwavefunct,crosscorr}.
One of the challenges is to prevent processes where pairs of entangled
particles reach the same lead, \emph{i.e.} are not spatially
separated.  Electrons from Cooper pairs are entangled in spin and
energy space, and separation of pairs into different leads using
ferromagnets or quantum dots has been suggested
\cite{entanglerfilters}. Upon filtering, only the spin or energy part
of the two-particle wave function collapses, depending on whether
ferromagnets or quantum dots are used. Respectively, energy or spin
entanglement remains \cite{entanglerwavefunct}. Here we consider
separation by ferromagnets.

Solid state entanglers have been analyzed in Refs.
\cite{NS-entanglers,entanglerfilters,entanglerwavefunct,crosscorr} in
terms of currents, noise and cross correlations. A more direct
approach, describing the elementary charge transfer processes in terms
of experimentally controllable parameters is certainly desirable. We
demonstrate how this is possible through the full distribution of
current fluctuations, the full counting statistics (FCS), of the solid
state entangler \cite{levitov:jetp58,belzig:NS-FCS,nazarov:qnoise}.
The FCS provides complete information about currents, noise, cross
correlations and higher cumulants, and even more importantly, allows
direct access to the probability for transfer of charge between
different parts of the device.

\begin{figure}[hh] 
\begin{picture}(0,0)%
\includegraphics{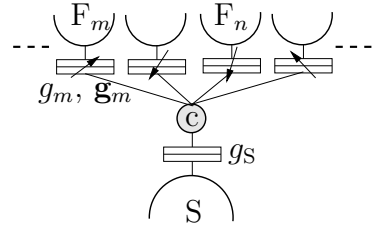}%
\end{picture}%
\setlength{\unitlength}{1184sp}%
\begingroup\makeatletter\ifx\SetFigFont\undefined%
\gdef\SetFigFont#1#2#3#4#5{%
  \reset@font\fontsize{#1}{#2pt}%
  \fontfamily{#3}\fontseries{#4}\fontshape{#5}%
  \selectfont}%
\fi\endgroup%
\begin{picture}(7888,4495)(-643,-6205)
\put(3151,-6136){\makebox(0,0)[lb]{\smash{{\SetFigFont{5}{6.0}{\familydefault}{\mddefault}{\updefault}{\color[rgb]{0,0,0}{\large S}}%
}}}}
\put(3151,-4111){\makebox(0,0)[lb]{\smash{{\SetFigFont{5}{6.0}{\familydefault}{\mddefault}{\updefault}{\color[rgb]{0,0,0}{\large c}}%
}}}}
\put(3751,-2011){\makebox(0,0)[lb]{\smash{{\SetFigFont{5}{6.0}{\familydefault}{\mddefault}{\updefault}{\color[rgb]{0,0,0}{\large F$_n$}}%
}}}}
\put(751,-2011){\makebox(0,0)[lb]{\smash{{\SetFigFont{5}{6.0}{\familydefault}{\mddefault}{\updefault}{\color[rgb]{0,0,0}{\large F$_m$}}%
}}}}
\put( 76,-3511){\makebox(0,0)[lb]{\smash{{\SetFigFont{5}{6.0}{\familydefault}{\mddefault}{\updefault}{\color[rgb]{0,0,0}{\large $g_m,\,\vec{g}_m$}}%
}}}}
\put(4051,-4786){\makebox(0,0)[lb]{\smash{{\SetFigFont{5}{6.0}{\familydefault}{\mddefault}{\updefault}{\color[rgb]{0,0,0}{\large $g_\text{S}$}}%
}}}}
\end{picture}%
\caption{Circuit theory representation of generic F-S
  entangler. Entangled electrons from the singlet superconductor
  S enter the cavity c through a tunnel barrier with
  conductance $g_\text{S}$ and escape through ferromagnetic interfaces
  with conductance $g_n$ and spin polarization $|\vec g_n|/g_n$ into
  drains F$_n$.  Arrows indicate magnetization directions $\vec
  g_n/|\vec g_n|$.
  \label{fig:ct}}
\end{figure}

We consider the singlet superconductor-ferromagnet (S-F) device shown
in Fig. \ref{fig:ct}. A normal metal cavity (c) is connected to one
superconducting terminal and several ferromagnetic terminals via
tunnel junctions. The cavity is under the influence of proximity
effect. In this device, charge transport occurs via two processes: i)
Transfer of Cooper pairs out of the superconductor by Andreev
reflection and ii) distribution of the entangled quasi-particles among
the ferromagnetic leads.  The distribution can occur via Direct
Andreev (DA) reflection, where a entangled pair is transferred into
lead F$_n$ or crossed Andreev reflection (CA), where each particle of
the entangled pair is transferred into spatially separated leads F$_m$
and F$_n$ ($n \ne m$). CA produces spatially separated entangled
electrons. Since the ferromagnetic terminals are at the same voltage
and we consider zero temperature, there is no direct electron
transport between the ferromagnetic terminals \cite{Falci:epl01}.

Our general results for the counting statistics show that the
processes i) and ii) are independent and therefore the statistics can
be factorized. This novel factorization and the probability
distribution for process ii) reveals the precise dependence of the
probabilities for CA and DA processes on the experimentally measurable
conductances and polarizations of the ferromagnetic leads. We find
that the probability that the electrons of a Cooper pair transferred
into the cavity are detected in terminals $m$ and $n$ is
\begin{equation}
  \label{eq:prob-normalized}
  p_{mn}=\left(g_mg_n-\vec g_m\vec g_n\right)/(g^2-\vec g^2)\,, 
\end{equation}
where $g=\sum_ng_n$ and $\vec g=\sum_n\vec g_n$. The probabilities
$p_{mn}$ depend solely on the conductances $g_n$ and spin polarization
conductances $\vec g_n$ of the ferromagnetic leads which can be
determined by magnetoresistance measurement in the normal
state. Eq.~(\ref{eq:prob-normalized}) shows that the detected
two-particle processes originate from a \emph{pure spin singlet}
density matrix subspace \cite{lorenzo:prl05}. We emphasize that
\eqref{eq:prob-normalized} in combination with the Cooper pair
transfer probability, to be discussed below, allows for an unambiguous
identification of all statistical properties of the charge transfer.
Using the magnetization dependence of the probabilities one can
violate Bell's inequality straightforwardly and demonstrate
entanglement. In a device with \textit{e.g.} two ferromagnetic
terminals, the probability to separate the entangled quasi-particle
pair into different leads is enhanced in an antiparallel magnetization
configuration.

It is quite remarkable that all statistics (noise, cross correlations,
and higher order cumulants) in this device are completely determined
by the \emph{normal state} transport conductances between the cavity
and the terminals and no additional parameters need to be
introduced. The relative orientations of the magnetizations are the
control parameters in an experimental situation. The fraction of the
CA current and, therefore, the spatially separated entangled pair
currents follow from these. 

Ferromagnet detection of entangled spin singlets from a ballistic
normal conductor was considered in Ref.  \cite{lorenzo:prl05}.  In
that device, there are also one-particle transfers, which can
contribute substantially to the current and the noise.  Another
important qualitative difference between our device and the system in
Ref. \cite{lorenzo:prl05} is the latter's strong coupling to the
detectors which can distort the spin singlets emitted from the source
and induce triplet correlations upon detection.  Also, in the limit of
a weak coupling to the detectors, there are no two-particle processes
in the system of Ref.  \cite{lorenzo:prl05}. In contrast, the
electrons in our S-F device are always detected from a pure spin
singlet state.

The charge transfer probabilities are obtained by identifying the
elementary processes in the many-body charge counting statistics. The
statistics is determined by the cumulant generating function (CGF)
$S(\chi_1,\ldots,\chi_N)=S(\{\chi_n\})$ of the probability
$P\left(\{N_n\}\right)$ to transfer in a time interval $t_0$, $N_1$
electrons to $F_1$, $N_2$ electrons to $F_2$, and so on. Our main
finding is the statistics
\begin{subequations}
\begin{eqnarray}
  \label{eq:prob1}
  P\left(\{N_n\}\right) & \equiv & \int \frac{d^n\chi}{(2\pi)^n} \exp^{S(\{\chi_n\})-i\sum_n\chi_nN_n}\\
   \label{eq:prob2}
& = & P_\text{S}\left(\sum\nolimits_nN_n\right) P\left(\{N_n\} \middle | \sum\nolimits_nN_n\right)
\end{eqnarray}
\label{eq:probs}
\end{subequations}
for $\sum_nN_n$ even and positive. The interpretation of this result
is that the charge transfer is given by two independent 
processes. The first factor $P_S(2N)$ is the probability that
$N=\sum_nN_n/2$ Cooper pairs are emitted from the superconducting
source terminal into any of the detectors. The second factor
$P\left(\{N_n\}|2N\right)$ in ~(\ref{eq:prob2}) is the conditional
probability that $N_n$ out of the $2N$ electrons have been transferred
into ferromagnetic terminal F$_n$. Below we will explain in detail how
our calculation yields concrete expressions for the elementary
processes described by $P\left(\{N_n\}|2N\right)$. These results
facilitates a unique interpretation of the transfer of spin singlet
electron pairs.

To complete the full statistical description, we now supply the
microscopic expressions for the two probabilities in \eqref{eq:prob2}.
The Cooper pair transfer probability is obtained from $P_\text{S}(2N)
= \int d\chi/(2\pi) \mathrm{exp}(S_\text{S}(\chi)-iN\chi)$ with a CGF
$S_\text{S}(\chi)$ given by
\begin{equation}
  \label{eq:prob-super}
    \frac{t_0 V}{\sqrt{2}e}\sqrt{g_\Sigma^2+\sqrt{ 
      (g_\text{S}^2-g^2+\vec g^2)^2 +4g_\text{S}^2(g^2-\vec g^2) e^{2i\chi}}}\,,
\end{equation}
where $g_\Sigma^2=g_\text{S}^2+g^2+\vec g^2$. The contact to the
superconducting terminal is characterized by a spin-independent
conductance $g_\text{S}$. The $\pi$-periodicity of $S_\text{S}(\chi)$
on $\chi$ ensures that an even number of charges is transferred. The
$2N$ electrons are distributed among the F$_n$ terminals according to
the multinomial distribution $P\left(\{N_n\}|2N\right)=\int d^n\chi/(2\pi)^n
\mathrm{exp}(S_\text{N}(\{\chi_n\})-i\sum_n\chi_nN_n)$ with a CGF
\begin{equation}
  \label{eq:prob-multi}
  S_\text{N}(\{\chi_n\}) = N \ln\left( \sum\nolimits_{mn} 
    p_{mn}e^{-i\chi_m-i\chi_n}\right)\,.
\end{equation}
The concrete form of the two-particle probabilities $p_{mn}$ to detect
one charge in terminal $m$ and one in terminal $n$ is given in
(\ref{eq:prob-normalized}). We will explain below how our calculation
based on the circuit theory of full counting statistics determines Eqs.\eqref{eq:prob-normalized},\eqref{eq:prob-super}, and \eqref{eq:prob-multi}.

We emphasize that our interpretation in terms of two independent
processes of charge transfer is based on a detailed calculation with a
general result for the counting statistics. We have not made any a
priori assumptions on the initial state of the superconducting source
terminal or the ferromagnetic terminals, except that they are
reservoirs at zero temperature with a voltage bias $eV$ applied. The
\emph{direct result} of our calculation is the CGF $S(\{\chi_n\})$ of
the S-F entangler of Fig. \ref{fig:ct}. Its explicit expression is
found in \eqref{eq:prob-super} by replacing the factor $\exp^{2i\chi}$
with exp$\{S_\text{N}(\{\chi_n\})/N\}$ given in (\ref{eq:prob-multi}).
The factorization in \eqref{eq:probs} can be proven straightforwardly
from $S(\{\chi_n\})$. Actually, such a factorization is valid for any
CGF where the $\chi$ dependence is exp$\{S_\text{N}(\{\chi_n\})/N\}$,
irrespectively of its form or the probabilities $p_{mn}$.

We now discuss some consequences of the charge counting statistics.
FCS enables us to express the current and noise correlations in a
compact and meaningful form. The currents $I_n=(ie/t_0)\partial
S(\{\chi_n\}) / \partial\chi_n |_{\chi_n=0}$ are
\begin{align}
  \label{eq:current}
  I=GV,~ G=\frac{g_\text{S}^2(g^2-\vec g^2)}{\sqrt{g_\text{S}^2+g^2}(g_\text{S}^2+g^2-\vec g^2)},~ I_n=Ip_n,
\end{align}
where $p_n=\sum_mp_{mn}$ is the probability to detect one of the
electrons in terminal $n$, irrespective where the second electron
goes. The combined probabilities can be directly accessed in the noise
correlators between current fluctuations in terminals $m$ and $n$,
$C_{mn}=(-2e^2/t_0)\partial^2 S(\{\chi_n\})
/ \partial\chi_m \partial\chi_n |_{\chi_{n,m}=0}$:
\begin{equation}
  \label{eq:noises}
  C_{mn}=2eI\left[p_{mn}+p_n\delta_{mn}-2(1-F_2)p_mp_n\right].
\end{equation}
The Fano factor for Cooper pair transport is defined as the ratio of
the full current noise $C=\sum_{mn}C_{mn}$ to the Poissonian noise of
doubled charges, $F_2=C/4eI$, and is explicitly found
$2(1-F_2)=[5-\vec g^2/(g_\text{S}^2+g^2)]x/(1+x)^2$ where
$x=g_\text{S}^2/(g^2-\vec g^2)$. These expressions for the current and
the noise provide a transparent interpretation of the transport
processes. The current in (\ref{eq:current}) is proportional to
$g_\text{S}^2(g^2-\vec g^2)$, since two particles have to tunnel
through the double junction to transfer a Cooper pair from S. The
denominator is due to the proximity effect
\cite{volkov:21,nazarov:prl94,morten:prb06} and enhances the current
drastically in comparison to calculations based on the tunneling
Hamiltonian \cite{Falci:epl01}. The current into each terminal $I_n$
is then weighted according to the probability $p_n$.  We might also
distinguish the contributions to the current originating from crossed
and direct Andreev reflection. The probability to detect a DA
reflection in terminal $n$ is given by $p_{nn}$, and the probability
for CA detection in different terminals $m\neq n$ is given by
$p_{mn}$.  We find the ratio of the crossed current to the total
current as
\begin{equation}
  \label{eq:curr_ca}
  \frac{I_n^\text{CA}}{I_n} = \frac{p_n-p_{nn}}{p_n} = 
  \frac{g_n(g-g_n)-\vec g_n(\vec g-\vec g_n)}{g_ng-\vec g_n\vec g}\,.
\end{equation}
This ratio is independent of the coupling to the
superconducting terminal.  We further observe that the crossed current
is enhanced by increasing the polarization of the contact $n$ and is
additionally favored by aligning the magnetization $\vec g_n$
opposite to the average magnetization $\vec g$. These results are a
direct consequence of the spin-singlet nature of the Cooper
pairs. Enhancing the magnitude of the polarization $|\vec g_n|/g_n$ of
one terminal reduces the total current, but enhances the crossed part
of the Andreev current, since the tunneling of one spin-singlet
electron-hole pair through the same contact is strongly
suppressed.

The sign of cross correlations in three-terminal beam splitters has
been considered for various devices both experimentally
\cite{oberholzer:046804} and theoretically
\cite{crosscorr,borlin:197001}.  Studies of noise
\cite{burkard:prb16303} and FCS \cite{taddei:prb02} for a beam
splitter with entangled electrons show that entanglement gives
qualitatively different noise characteristics compared to transport of
non-entangled electrons. The physical origin of positive and negative
contributions to the cross correlators can in our case be understood
from the dependence on the two-particle probabilities in
\eqref{eq:noises}. CA reflection leads to positive cross correlations
since two particles are transferred simultaneously into F$_m$ and
F$_n$ (bunching behavior) \cite{burkard:prb16303}. A negative
contribution (anti bunching) that does not depend on entanglement, is
induced by the fermion exclusion principle: The transfer of one
electron-hole pair into F$_{m}$ by DA reflection, prevents the
simultaneous transfer of another pair into F$_{n}$.  However, if the
electron-hole pair transfers are not temporally correlated (Poissonian
statistics), the exclusion principle does not affect the cross
correlations. This is the case when there is strong asymmetry in the
junction conductances $g$ and $g_\text{S}$ ($g \gg g_\text{s}$ or $g
\ll g_\text{s}$) so that the Fano factor $F_2=1$. In this limit the
negative contribution $-2(1-F_2)p_mp_n$ in \eqref{eq:noises} vanishes.
Scattering matrix calculations give similar results for $C_{mn}$
\cite{torres:epjb99}.

The strongly asymmetric case is particularly interesting since the
cross correlations ($m\neq n$) $C_{mn}=2eIp_{mn}$, are a
\textit{direct} measure of the probability that electrons from a
Cooper pair are transferred into different terminals.

\begin{figure}[hh]
  \includegraphics[scale=0.7]{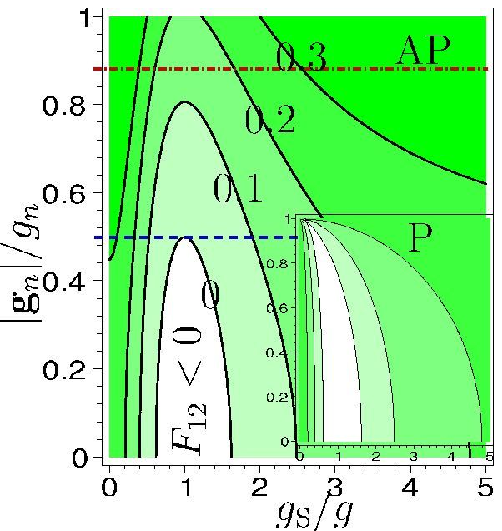}
  \includegraphics[scale=0.44]{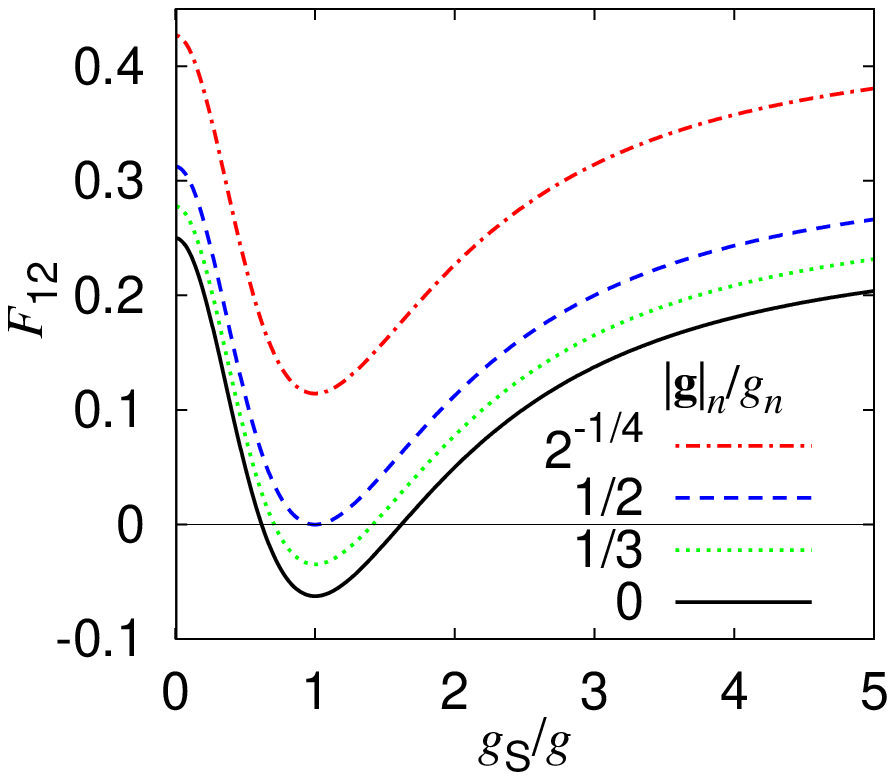}
  \caption{(color online) Left: Contour plots of $F_{12}$ for
    antiparallel and parallel (inset) configuration ($|\vec g_1|=|\vec
    g_2|$). Positive regions in green. Right: Plots of $F_{12}$ in
    antiparallel configuration as a function of the conductance
    asymmetry $g_\text{S}/g$, legend denotes the value of the
    polarization $|\vec g_n|/g_n$.  Red and blue horizontal lines in
    left panel correspond to red and blue curves in right panel.}
    \label{fig:plots}
\end{figure}

To illustrate our theory, let us now consider the three-terminal
version of Fig.~\ref{fig:ct} with the superconducting source terminal
$S$ and two ferromagnetic drains F$_1$ and F$_2$. The ferromagnetic
magnetizations can in this device be utilized as filters to produce
currents of entangled electrons in separated leads. Let us consider
$|\vec g_1|=|\vec g_2|$ in the following and define Fano factors
$F_{mn}=C_{mn}/(2eI)$. The autocorrelation noise $F_{11(22)}$ will be
reduced in antiparallel alignment $\vec g_1=-\vec g_2$ as compared to
a S-N system ($\vec g_n=0$) due to enhancement of CA. The cross
correlation $F_{12}$, shown in Fig. \ref{fig:plots} can have both
positive and negative sign depending on the conductance asymmetry
$g_\text{S}/g$ and the spin polarization. The positive contribution to
$F_{12}$ is proportional to $g_1g_2+(-)|\vec g_1|\,|\vec g_2|$ in the
parallel (antiparallel) alignment demonstrating how spin filtering of
entangled pairs enhances (reduces) the correlation between currents in
F$_1$ and F$_2$ with respect to an S-N system \cite{crosscorr}. Note
that for sufficiently large spin polarization, $F_{12}$ can be
positive for the entire range of $g_\text{S}/g$ in the antiparallel
alignment (region above blue line in left panel of Fig.
\ref{fig:plots}), whereas it remains always negative in the parallel
alignment for $g_\text{S}/g \simeq 1$ (inset of Fig.
\ref{fig:plots}). The change of sign in $F_{12}$ by switching from
antiparallel to parallel alignment is due to the enhanced probability
of CA events, see \eqref{eq:curr_ca}.

We will finally outline the calculation that yields the FCS of the
considered devices. We utilize the circuit theory of mesoscopic
superconductivity \cite{nazarov:sm99,belzig:NS-FCS} and represent the
circuit in terms of terminals, cavities and connectors. Terminals are
described by equilibrium quasiclassical Green's function matrices
$\check{G}_n$ determined by electrochemical potential and temperature.
Our notation for matrix subspace is: $\bar{}$ for spin, $\hat{}$ for
Nambu, and $\check{}$ for Keldysh.  Pauli matrices are denoted
$\tau_j$. At zero temperature we consider $0<E\leq eV$ where the
Green's functions for all ferromagnetic terminals F$_n$ are $\check{G}
= \hat{\tau}_3\check{\tau}_3 + (\check{\tau}_1+i\check{\tau}_2)$ where
$V$ is the voltage of the ferromagnetic terminals and $E$ the
quasiparticle energy. The superconductor S is at zero voltage and has
Green's function $\check{G}_\text{S} = \hat{\tau}_1$, where we assume
$E\ll \Delta$, $\Delta$ being the gap of S. The terminals are
connected to a cavity c which is under the influence of proximity
effect from S. The cavity is described by an unknown Green's function
$\check{G}_\text{c}$, assumed isotropic due to chaotic or diffusive
scattering. We assume that c is large enough so that charging effects
can be neglected, and small enough so that $\check{G}_\text{c}$ is
spatially homogeneous. The circuit theory is formulated in terms of
generalized matrix currents $\check{I}_j$ in
spin$\otimes$Nambu$\otimes$Keldysh matrix space and from the matrix
current conservation $\sum_j\check{I}_j=0$. This determines the Greens
function on the node together with the normalization condition $\check
G_c^2=1$. The matrix currents can have arbitrary structure, and allow
to derive the FCS by introducing the counting fields $\chi_n$ for each
terminal according to \cite{belzig:NS-FCS} $\check{G}_n(\chi)=
\exp^{i\chi_n\hat{\tau}_3\check{\tau}_1/2}\check{G}
\exp^{-i\chi_n\hat{\tau}_3\check{\tau}_1/2}$. Spin active connectors
are taken into account by spin dependent transmission and reflection
amplitudes $t_{k,\sigma}^n$ and $r_{k,\sigma}^n$ for particles
incident on the interface $n$ from the cavity side in channel $k$ with
spin $\sigma$. The matrix current through a spin active tunnel barrier
between c and F$_n$ evaluated at the cavity side is
\cite{huertas:misc,BC-applications} $\check{I}_n = \left[g_n
  \check{G}_n/2 +
  \left\{\vec{g}_n\bar{\boldsymbol{\tau}}\hat{\tau}_3,\check{G}_n\right\}/4,\check{G}_\text{c}\right]$.
Here, $g_n=g_\text{Q}\sum_{k,\sigma}|t_{k,\sigma}^{n}|^2$ is the
tunnel conductance and $g_\text{Q}=e^2/h$ the conductance quantum. The
magnetization direction is encoded in the direction of $\vec{g}_n$,
and the conductance polarization in that quantization axis is $|\vec
g_{n}| = g_\text{Q} \sum_k (|t^{n}_{k,\uparrow}|^2 -
|t^{n}_{k,\downarrow}|^2)$. We have neglected here an additional term
related to spin dependent phase shifts upon reflection at the
interface \cite{huertas:misc,BC-applications}, as these are small in
some material combinations or can be suppressed by a thin,
non-magnetic oxide layer \cite{Tedrow:prl86}. The matrix current
between c and S is
$\check{I}_\text{S}=g_\text{S}[\check{G}_\text{S},\check{G}_\text{c}]/2$
\cite{nazarov:sm99}.  We take into account the spin structure of
matrix currents $\check{I}_n$ and Green's functions in S-F-systems,
and derive the CGF in the linear response regime and for $eV\ll
\Delta$, generalizing Ref.~\cite{borlin:197001}: $S=t_0/(4e^2)\int\d E
\sum_p\sqrt{\lambda_p^2},$ where $\{\lambda_p\}$ is the set of
eigenvalues of the matrix $\check{M}$ defined by writing matrix
current conservation in the cavity $\sum_j\check{I}_j\equiv
[\check{M},\check{G}_\text{c}]=0$. The non-trivial spin matrix
structure of $\check{I}_n$ determines the magnetization dependence of
transport processes in the system. Carrying out this procedure yields
the FCS for the setup in Fig.~1.

In conclusion, we have investigated the elementary charge transfer
processes of a S-F entangler. Charge transfers occurs via two
statistically independent processes, i) Cooper pairs are transferred
out of the superconductor by Andreev reflection and ii) entangled
quasiparticles are distributed among the different ferromagnetic
leads. The probabilities for entangled electrons to flow into
spatially separated leads are completely determined by experimentally
measurable conductances and polarizations. This allows complete
knowledge of the statistics of charge transfer in the S-F entangler.

\begin{acknowledgments}
  We would like to thank Yu. V. Nazarov, A. Di Lorenzo, A.A. Golubov,
  A. Brinkman, D. Sanchez, and M. B\"{u}ttiker for discussions. This
  work was supported in part by The Research Council of Norway through
  Grants No. 167498/V30, 162742/V00, 1534581/432, 1585181/143,
  1585471/431, the DFG through SFB 513, the Landesstiftung
  Baden-W\"{u}rttemberg, the NSF Grant No.  PHY99-07949, and EU via
  project NMP2-CT-2003-505587 'SFINx'.
\end{acknowledgments}

\bibliography{/home/gudrun/janpette/artikkel/fs.bib}

\end{document}